\documentclass[aps,pra,twocolumn,bibnotes]{revtex4}
\usepackage{graphicx}
\usepackage{graphics}
\begin{document}

\title{Erratum : Squeezing and entanglement delay using slow light}

\author{G.~H\'etet } 
\author{A.~Peng}
\author{M.~T.~Johnsson}
\author{ M.~T.~L.~Hsu}
\author{O.~Gl\"ockl}
\author{P.~K.~Lam}
\email[Email: ]{ping.lam@anu.edu.au}
\author{H~.A.~Bachor}
\author{J.~J.~Hope}
\affiliation{ARC COE for Quantum-Atom Optics, Australian National University, Canberra, ACT 0200, Australia}

\date{\today}

\begin{abstract}
An inconsistency was found in the equations used to calculate the variance of the quadrature fluctuations of a field propagating through a medium demonstrating electromagnetically induced transparency (EIT).  The decoherence term used in our original paper introduces inconsistency under weak probe approximation.  In this erratum we give the Bloch equations with the correct dephasing terms.  The conclusions of the original paper remain the same.  Both entanglement and squeezing can be delayed and preserved using EIT without adding noise when the decoherence rate is small $ \gamma \gamma_{bc}' {/} \Omega_{c} \ll 1$.
\end{abstract}


\maketitle

In our original paper \cite{peng}, we intended to describe a model where decoherence between the two ground states arises only from off-diagonal terms in the atomic density matrix. That is, the dephasing was supposed to be due to effects such as elastic collisions that cannot change the populations of the two ground states, only their relative phase. This type of dephasing is the dominant source of loss in current experiments in vapor cells \cite{Lvov}. In our Bloch equation, we instead included dephasing terms that allowed population transfer between the two ground states due to inelastic collisions between atoms.
In this erratum we show that allowing these population transfer terms leads to an inconsistency in the equations derived under the weak probe approximation. We then modify the Bloch equations in our original paper to use off-diagonal dephasing only, and recalculate the noise added to the field. If one solves for $\langle\sigma_{ba}\rangle$ using Eq.(4) and (5) of our original paper, and substitutes back into the equations of motion for the populations, the weak probe approximation yields 
\begin{eqnarray}
\langle\sigma_{bb}\rangle=\frac{-2 g^{2}}{\gamma_{ba}\gamma_{bc}+|\Omega_{c}^{2}|}|\langle\hat{\mathcal{E}}\rangle|^{2}.
\end{eqnarray}
This relation contradicts our weak probe approximation since it assumes that almost all the population is in state $|b\rangle$ throughout the process. 
The spectral variance of the output field is

\begin{eqnarray}
\lefteqn{ \mathcal{S}_{out}(\omega)   = \mathcal{S}_{in}(\omega) e^{-2
\Re\{ \Lambda(\omega) \} L}  \label{outspectrum1} }\nonumber \\
 & & + (1-e^{-2\Re\{ \Lambda(\omega) \} L} ) \left( 1-\frac{\gamma_{bc}(\omega^{2}+\gamma_{bc}^{2})}{\gamma(\omega^{2}+\gamma_{bc}^{2})+\gamma_{bc}|\Omega_{c}|^{2}} \right), \nonumber \\
\end{eqnarray}
which violates the canonical commutation relation.


We now solve the Heisenberg-Bloch equations using off-diagonal dephasing only, with the dephasing rate denoted by $\gamma_{bc}'$. The revised Heisenberg-Bloch equations of motion are
\begin{eqnarray}
\dot{\hat{\sigma}}_{bb} & = & \gamma_b \hat{\sigma}_{aa} - i g
\hat{\mathcal{E}} \hat{\sigma}_{ab} + i g^{\ast}
\hat{\mathcal{E}}^{\dagger} \hat{\sigma}_{ba} + \hat{F}_{bb} \nonumber \\
\dot{\hat{\sigma}}_{cc} & = & \gamma_c \hat{\sigma}_{aa} - i \Omega_c
\hat{\sigma}_{ac} + i \Omega_c^{\ast} \hat{\sigma}_{ca} +
\hat{F}_{cc} \nonumber \\
\dot{\hat{\sigma}}_{ba} & = & -\gamma_{ba} \hat{\sigma}_{ba} + i g
\hat{\mathcal{E}} (\hat{\sigma}_{bb} - \hat{\sigma}_{aa} ) + i
\Omega_c \hat{\sigma}_{bc} + \hat{F}_{ba} \nonumber \\
\dot{\hat{\sigma}}_{bc} & = & -\gamma_{bc}' \hat{\sigma}_{bc} - i g
\hat{\mathcal{E}} \hat{\sigma}_{ac} + i \Omega_c^{\ast}
\hat{\sigma}_{ba} + \hat{F}_{bc} \nonumber \\
\dot{\hat{\sigma}}_{ac} & = & -\gamma_{ac} \hat{\sigma}_{ac} - i
g^{\ast} \hat{\mathcal{E}}^{\dagger} \hat{\sigma}_{bc} + i
\Omega_c^{\ast} (\hat{\sigma}_{aa} - \hat{\sigma}_{cc}) +
\hat{F}_{ac} \nonumber \\
\left( \frac{\partial}{\partial t} \right. & + & \left. c
\frac{\partial}{\partial z} \right) \hat{\mathcal{E}}  =  i
g^{\ast} N \hat{\sigma}_{ba} \label{motion2}
\end{eqnarray}

Following the same approach as the original paper we can obtain a set of three closed
equations under the weak probe approximation.
Substituting  $\langle\sigma_{ba}\rangle$ into the equations of motion for the population equations in the steady state regime, we find that the equations are now valid to second order in the weak probe approximation. When calculating the variance of the output field we find 
\begin{eqnarray}
\mathcal{S}_{out}(\omega)  & = & \mathcal{S}_{in}(\omega) e^{-2\Re\{ \Lambda(\omega) \} L}  \label{outspectrum2} \nonumber \\
&  & + (1-e^{-2\Re\{ \Lambda(\omega) \} L} ). 
\end{eqnarray}

Our conclusion is similar to our previous paper. This treatment of the propagation of light through a three level system shows that no additional noise is introduced into the light field beyond that which is necessary to preserve the canonical commutation relation of the field at the output.


\end{document}